\documentclass[preprint,aps,floats,showpacs]{revtex4}
\usepackage{graphicx}
\usepackage[dvips]{pstcol}
\def\upper{\hbox{${}^{\perp}$}\!}
\def\uppar{\hbox{${}^{\parallel}$}\!}

\definecolor{darkgreen}{rgb}{.1,.4,.1}

\begin{document}
\newcommand{\beq}{\begin{eqnarray}}
\newcommand{\eeq}{\end{eqnarray}}

\title{Bouncing pre-big bang on the brane}

\author{Stefano Foffa}
\affiliation{
D\'epartment de Physique Th\'eorique, Universit\'e de Gen\`eve,
\\24, quai Ernest-Ansermet, CH-1211 Gen\`eve 4, Switzerland
\\email: foffa@amorgos.unige.ch}
\begin{abstract}
A regular bouncing universe is obtained in the context of a
dilaton-gravity brane world scenario. The scale factor starts in a
contracting inflationary phase both in the Einstein and in the string
frame, it then undergoes a bounce (due to interaction with the bulk
Weyl tensor), and subsequently enters into a decelerated expanding
era. This {\it graceful exit} is obtained at low curvature and low
coupling, and without violating the Null Energy Condition.
\end{abstract}

\maketitle

\section{Introduction}
According to the singularity theorems \cite{sing}, it is impossible to
avoid the big bang singularity within the context of General
Relativity, at least if some ``reasonable energy conditions'' hold.
In particular, the Einstein equations tell us that in order to have a
{\it bounce} (namely a transition from a contracting to an expanding
phase) in a flat or open universe
\footnote{In the case of a closed universe it is possible to find
bouncing solutions without violating the NEC. However, such
solutions are somehow pathological, being them unstable, or requiring some
kind of fine tuning, see for example \cite{curv}; in the present treatment
curvature-driven bouncing solutions will not be considered.}, the
so-called Null Energy Condition (NEC) (which for a perfect fluid
states $\rho+p\geq0$) must be violated.

A violation of the NEC is usually considered unacceptable in that it
generally implies a violation of causality, but this is not always the
case. For example, in various extended theories of gravity it is
possible that terms coming from modifications of General Relativity
(e.g. close to the Planck scale) give an (apparent, and thus not
troublesome) violation of the NEC if they are interpreted as sources
in the modified Einstein equations; consequently, they can also drive
the evolution of the scale factor trough a bounce.
In such cases the main theoretical problem does not come from the
apparent violation of the NEC, but from our ignorance on the exact
nature and form of the new terms; this fact casts some doubt on
the reliability of the bouncing solutions that can be obtained in this
way.

This is exactly what happens in the pre-big bang model \cite{prebig},
where the connection of the pre- to the post-big bang phase
(henceforth, the {\it graceful exit}) can be actually mapped into a
bounce if one uses Einstein frame variables, and thus requires the use
of NEC-violating terms \cite{exitprob}.
Such terms are already incorporated in the
model (they are loop and high curvature corrections dictated by string
theory, and the violation of the NEC by means of such effective
sources is considered to be acceptable), but our knowledge of their
exact form is far from being complete.
As a consequence, all the realizations of the graceful exit presented
so far \cite{exits} can not be taken as completely reliable in that
they imply some assumption on the exact form of these loop and high
curvature corrections
\footnote{In general Brans-Dicke models the situation is more
variegated and a bouncing behavior can be obtained in some cases, see
\cite{PPeter}.}.

In the brane world scenario the situation is quite different: due to
the presence of interaction terms between the brane and the bulk, a
bounce can quite easily be obtained without any NEC violation
and without the need of any quantum gravity or high curvature
effect. This has been obtained for example in \cite{MuhPel} by means
of a slight modification of the Randall-Sundrum model \cite{RaSu}
(for other considerations on bouncing brane world models, see
also \cite{MedKam}).

Generally speaking, one expects a bouncing universe in the brane world
scenario to be less difficult to obtain because it does not require
a dramatic change in the structure of bulk spacetime.
In other words, since in the Randall-Sundrum picture the singularity 
can be avoided by a timelike trajectory, there is no need to introduce
NEC violating sources in order to eliminate it (as is needed in
standard cosmology, where the big bang singularity is spacelike):
rather, it is sufficient to prevent the brane trajectory from reaching
it. So, in some sense, the singularity is still there in the bulk
(for which the singularity theorems are, after all, still valid),
but simply a brane observer never comes close to it.

This paper is organized as follows: in section \ref{II}, after having
discussed the general conditions for a bounce in a brane world
model, I will apply these considerations to a gravi-dilaton scenario,
where an explicit solution will be found in the Einstein frame.
Brane models in the presence of a scalar field have already been studied
in \cite{ChamRe}, \cite{brdil} and \cite{foffa}; in particular some
bouncing solutions have already been found in \cite{ChamRe} for a
domain wall in the presence of a Liouville potential.
In the present treatment, which extends the work of \cite{ChamRe} in that
it is valid for a generic matter content on the brane,
the emphasis will be put on the string frame behavior: in fact in
section \ref{III} the bouncing solution will be transformed in the string
frame and it will be compared with what one obtains in the pre-big
bang model,
whose basic features will also be briefly recalled. It will be shown
that such a solution can be seen as a realization of a {\it graceful
exit} in a pre-big bang model, and that this exit can be realized
without any use of quantum or high curvature corrections, and is thus
under complete control.
This is a concrete realization of the idea (suggested in
\cite{foffa}) that in brane world models there is room for dealing
with problems which are difficult or intractable in more conventional
scenarios.

A few comments and remarks will then conclude the paper.

\section{Bouncing universe in a brane world scenario}\label{II}
\subsection{General conditions for a bounce}\label{genweyl}
Following \cite{Shiro}, we write the projected Einstein equations
on the brane as
\beq\label{einbr}
G_{\mu\nu}&=& 8\pi G_{N}\tau_{\mu\nu}+
\frac{d-1}{d}{\kappa}^2\left[\uppar{T}_{\nu\mu} +
\frac{1}{d+1}\left(d \upper{T} - \uppar{T}\right)h_{\mu\nu}\right]
-\Lambda_{d}h_{\mu\nu}+\kappa^4 \pi_{\mu\nu}-E_{\mu\nu}\, ,
\eeq
where $\tau_{\mu\nu}$ is the energy momentum tensor on the brane,
$\uppar{T}_{\mu\nu}$ and $\upper{T}$ are, respectively, the
projections parallel and perpendicular to the brane of the bulk energy
momentum tensor\footnote{In the Randall-Sundrum model only a
cosmological constant is allowed to be present in the bulk, according
to the prescription that the matter fields should be confined on the
brane. Since in this work we are going to analyze a gravi-dilaton
system, the dilaton field is also allowed to propagate in the bulk,
and $T_{\mu\nu}$ is meant to represent the bulk energy momentum tensor
related to this field.}, $\Lambda_{d}$ the effective cosmological
constant on the brane, $E_{\mu\nu}$ the projection on the brane of the
bulk Weyl tensor, $h_{\mu\nu}$ the induced metric on the
$(d+1)$-brane, $\kappa$ the bulk gravitational coupling, $G_{N}$ the
effective brane Newton constant and, finally, $\pi_{\mu\nu}$ a term
quadratic in $\tau_{\mu\nu}$ given by the following expression:
\beq\label{pimunu}
\pi_{\mu\nu}&=&
-\frac{1}{4}\tau_{\alpha\mu}\tau^{\alpha}_{\nu}+
\frac{1}{4d}\tau\tau_{\mu\nu}+
\frac{1}{8}\tau^{\alpha\beta}\tau_{\alpha\beta}
h_{\mu\nu}-\frac{1}{8d}\tau^2 h_{\mu\nu}\, .
\eeq
By making a flat FRW ansatz for $h_{\mu\nu}$ (as already stated,
curvature-driven bounces will not be considered here), and a perfect
fluid one for the energy momentum tensors
$\tau^{\mu}_{\nu}\equiv\left(-\rho,\vec{p}\right)$,
$\uppar{T}^{\mu}_{\nu}\equiv\left(-R,\vec{P}\right)$,
one can write an equation for the time derivative of the Hubble
parameter on the brane:
\beq\label{hdot}
\dot{H}&=&-\frac{1}{d}\kappa^2 \left(R+P\right)
- \left(\frac{8\pi G_{N}}{d-1}+\frac{\kappa^4}{4 d}\rho\right)
\left(\rho + p\right)-\frac{d+1}{d(d-1)}E^{0}_{0}\, .
\eeq
As can be read in this equation, if the Weyl tensor is set to zero, it
is impossible to have $\dot{H}>0$, a necessary condition for a bounce,
without violating the NEC either in the bulk or on the brane,
or without having $\rho<0$ (this last condition deriving from
the term $\rho \left(\rho+p\right)$, which comes from the explicit
expression of $\pi_{\mu\nu}$ in terms of $\rho$ and $p$).
Thus, if one wants to realize a bounce without resorting to
``exotic'' forms of matter, one has to consider the case of a
nonvanishing projected Weyl tensor. More precisely:
\beq\label{weyl}
E^{0}_{0}<0\quad\quad\longrightarrow\quad\quad\dot{H}>0\, .
\eeq
The bouncing solution presented in \cite{MuhPel} exploits exactly this
kind of mechanism.

\subsection{Gravi-dilaton system and relevant equations}
With the previous condition in mind, let's consider the following
gravi-dilaton system in the Einstein frame
\beq\label{action}
S&=&\frac{1}{\kappa^2}\int{\rm d}^{d+2}x \sqrt{-g}\left[{\cal R}-
\frac{1}{2}\left(\nabla\Phi\right)^2 -
2\Lambda e^{\gamma\Phi}\right]+\nonumber\\
&&-\int{\rm d}^{d+1}y
\left[\sqrt{-h}\left(\frac{2 K^{\pm}}{\kappa^2} + \lambda
  e^{\alpha\Phi}\right)-
{\cal L}_{m}\left(\chi_{m}; \tilde{h}_{\mu\nu}\right)\right]\, ,
\eeq
where the brane matter fields $\chi_{m}$ are coupled to the dilaton through
the conformally rescaled brane metric  $\tilde{h}_{\mu\nu}$, defined as
\beq\label{htilde}
\tilde{h}_{\mu\nu}=e^{2\xi \Phi}h_{\mu\nu}\, .
\eeq

The bulk equations of motion and junction conditions are immediately
derived ($Z_{2}$ symmetry is assumed):
\beq\label{bulk}
\left\{
\begin{array}{l}
{\cal R}_{\mu\nu}=\frac{1}{2}{\nabla}_{\mu}\Phi{\nabla}_{\nu}\Phi+
\frac{2}{d}g_{\mu\nu}\Lambda e^{\gamma\Phi}\\
\nabla^2 \Phi=2\Lambda \gamma e^{\gamma\Phi}
\quad\quad\quad\quad\quad\quad\, ,
\end{array}\right.\hspace{2.1cm}
\eeq
\beq\label{junct}
\left\{
\begin{array}{l}
K_{\mu\nu}=-\frac{\kappa^2}{4}\left[\tau_{\mu\nu}-\frac{1}{d}
\left(\tau-\lambda e^{\alpha\Phi}\right)h_{\mu\nu}\right]\\
n\cdot\nabla\Phi=\frac{\kappa^2}{2}\left(-\xi\tau +
\alpha\lambda e^{\alpha\Phi}\right)
\quad\quad\quad\quad\, ,
\end{array}\right.\quad
\eeq
being $n_{\mu}$ the normal to the brane pointing {\em into} the bulk,
$K_{\mu\nu}\equiv\nabla_{\mu}n_{\nu}$, and $\tau_{\mu\nu}\equiv
-\frac{2}{\sqrt{-h}}\frac{\delta {\cal L}_{m}}{\delta h^{\mu\nu}}$.

If one makes the following static ansatz for the bulk metric
\beq\label{bulkmet}
{\rm d}s^2_{bulk}
=- F(r)^2{\rm d}T^2 +B(r)^2 {\rm d}r^2 +
A(r)^2{\rm d}\vec{x}^2_{d}\, ,
\eeq
which implies
\beq\label{brmet}
{\rm d}s^2_{brane}
=-{\rm d}t^2 + A[r(t)]^2{\rm d}\vec{x}^2_{d}\, ,
\eeq
then the junction conditions can be explicitly written as
\beq\label{jcexp}
&&H^2\equiv\left(\frac{\dot{A}}{A}\right)^2
=\frac{\kappa^4}{16 d^2}(\rho+\rho_{\lambda})^2
-\left(\frac{A'}{AB}\right)^2\, ,\\
\label{rho}&&\dot{\rho}+ H \left[d (1 + w)\rho +
\alpha\left(\Phi'\frac{A}{A'}\right)\rho_{\lambda} 
+\frac{1}{2 d}\left(\Phi'\frac{A}{A'}\right)^2
(\rho + \rho_{\lambda})\right]=0\, ,\\
\label{phi}&&(\rho + \rho_{\lambda})\Phi'\left(\frac{A}{A'}\right)=
2 d\xi\rho\left(d w -1\right) - 2 d \alpha\rho_{\lambda}\, ,
\eeq
where $\dot{A}\equiv\frac{d A}{d t}$, ${A'}\equiv\frac{\partial A}
{\partial r}$, $w\equiv\frac{p}{\rho}$,
and $\rho_{\lambda}\equiv\lambda e^{\alpha\Phi}$. 
The first two equations, coming from the junction conditions for
$K_{\mu\nu}$, can be seen as the effective Friedmann equation and
energy conservation equation on the brane (note the anomalous terms
containing $\Phi'$ in the latter, coming from the interactions of
the dilaton with the brane),
while the last one is the junction condition for the dilaton. 

\subsection{Bulk solution and brane equation}
It is now sufficient to know an explicit bulk
solution and to plug it in eqns.~(\ref{jcexp}), (\ref{rho}),
(\ref{phi}) in order to have an explicit set of equations for the
evolution of the brane.
All the static bulk solutions with the symmetry given by
eq.~(\ref{bulkmet}) are known (see \cite{charm});
here only the following simple class of solutions if considered:
\beq\label{charm}
\left\{
\begin{array}{l}
{\rm d}s^2_{bulk}
=-h(r)r^{s+\frac{2}{d}}{\rm d}t^2
+\frac{r^{s+\frac{2}{d}}}{h(r)}{\rm d}r^2 +
r^{\frac{2}{d}}{\rm d}\vec{x}^2_{d}\\
\quad\left[h(r)=\left(\frac{2\Lambda}{s}r^{-s}
e^{\gamma\Phi_0}+C\right),
s=\frac{\gamma^2}{2}-\frac{d+1}{d}\right]\, ,\\
~\\
\Phi=\Phi_0 - \gamma\log{r}\, .
\end{array}\right.
\eeq
Particularly relevant for the following will be the role of the
integration constant $C$, which can be shown to be zero {\it iff} the
bulk Weyl tensor is vanishing. 

Now one has to plug the bulk solution into the explicit junction
conditions. First of all, eq.~(\ref{phi}) gives two constraints on the
parameters of the model, namely
\beq\label{const}
\gamma=-2\xi(d w -1)\, ,\quad\quad\quad\quad \alpha=\frac{\gamma}{2}\, .
\eeq
Then, since the dilaton behavior is known, the energy conservation
equation (\ref{rho}) can be used to find out how the energy
density depends on $r$, and thus on $A(r)=r^{\frac{1}{d}}$: one finds
\beq\label{rhodia}
\rho=R_0 A^{- d \left(1 + w + \frac{\gamma^2}{2}\right)}\, ,
\eeq
with $R_0$ an integration constant. A similar equation for
$\rho_{\lambda}$ follows trivially from its definition in terms of
the dilaton.

Finally, everything can be plugged in the equation for $H$, which
one can rewrite as
\beq\label{acca}
H^2=\frac{\kappa^4}{16 d^2}\left(\rho^2+
2\rho_{\lambda}\rho+\rho_{\lambda}^2\right)
-\frac{1}{d^2} \left[\frac{2\Lambda}{s}e^{\gamma\Phi_0}
A^{- d \gamma^2} + C A^{- d \frac{\gamma^2}{2} - d - 1}
\right]\, .
\eeq
By taking into account of eq.~(\ref{rhodia}),
one can see that without the contribution of the Weyl
tensor, it is impossible to obtain a bouncing solution with
$w\geq-1$ (as is implied by the NEC and by the positivity $\rho$).
In fact, under these assumptions, the first term on the r.h.s.~of
(\ref{acca}) turns out always to be dominant over the others
for small $A$, and consequently
the solution can not be driven towards a bounce.
Thus the only strategy for getting a bouncing solution is to consider
the action of the last term, the one related to the Weyl tensor.

Moreover, the definition of $\rho_{\lambda}$ in terms of $\Phi$
implies that $\rho_{\lambda}^2$ has exactly the same $A$-dependence of
the first term in the squared brackets in eq.~(\ref{acca}). One can
take advantage of this fact by requiring an exact cancellation between
these two terms in the effective Friedmann equation,
cancellation that is realized
if $\frac{\kappa^4}{32}\lambda^2=\frac{\Lambda}{s}$. This condition
is the counterpart of the fine tuning required in the Randall-Sundrum
model between the brane tension and the bulk cosmological constant
in order to have a vanishing effective brane cosmological constant.

\subsection{Getting the bounce}
\begin{figure}
\begin{center}
\includegraphics[width=3.2in, angle=-0]{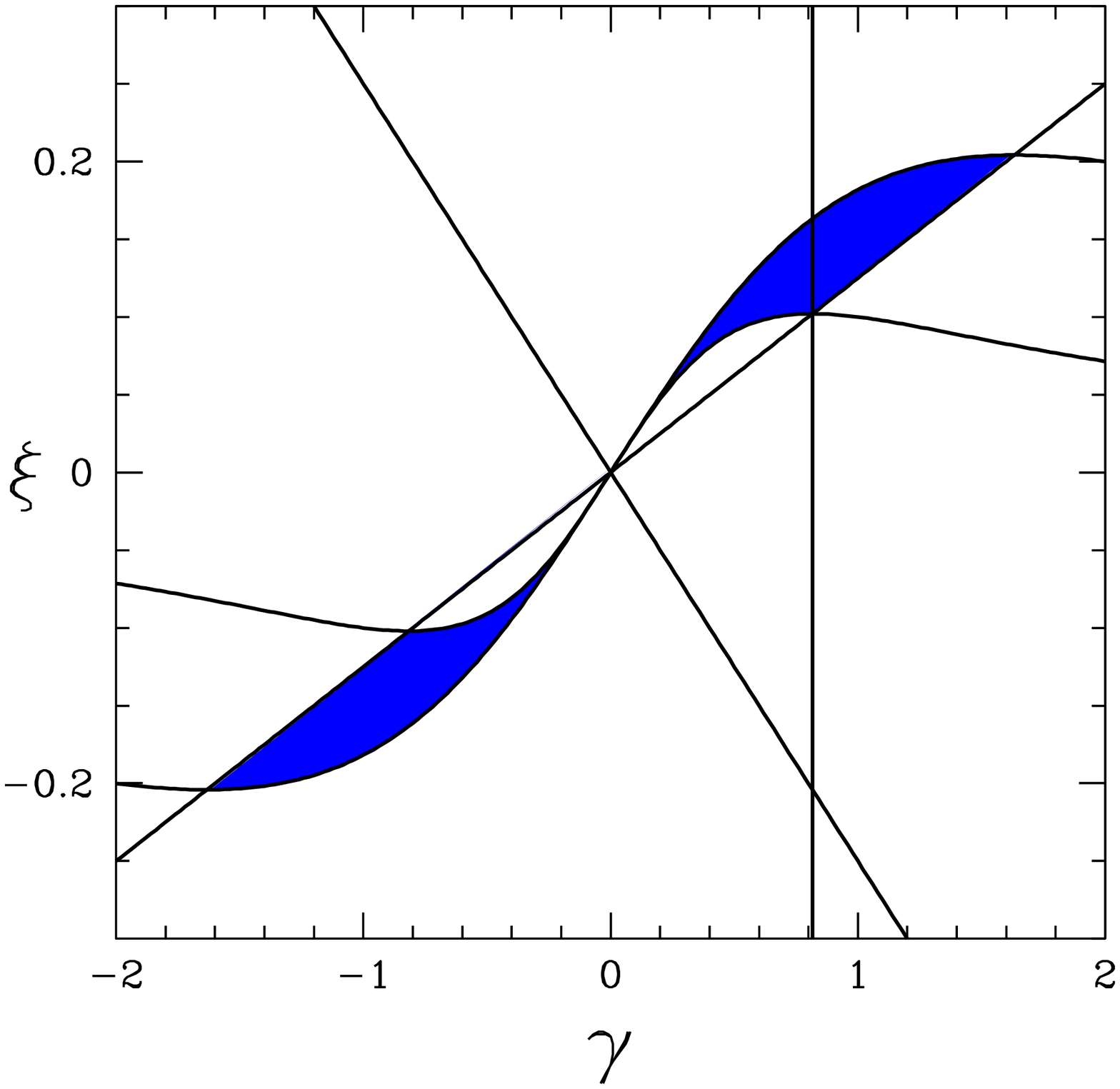}
\includegraphics[width=3.2in, angle=-0]{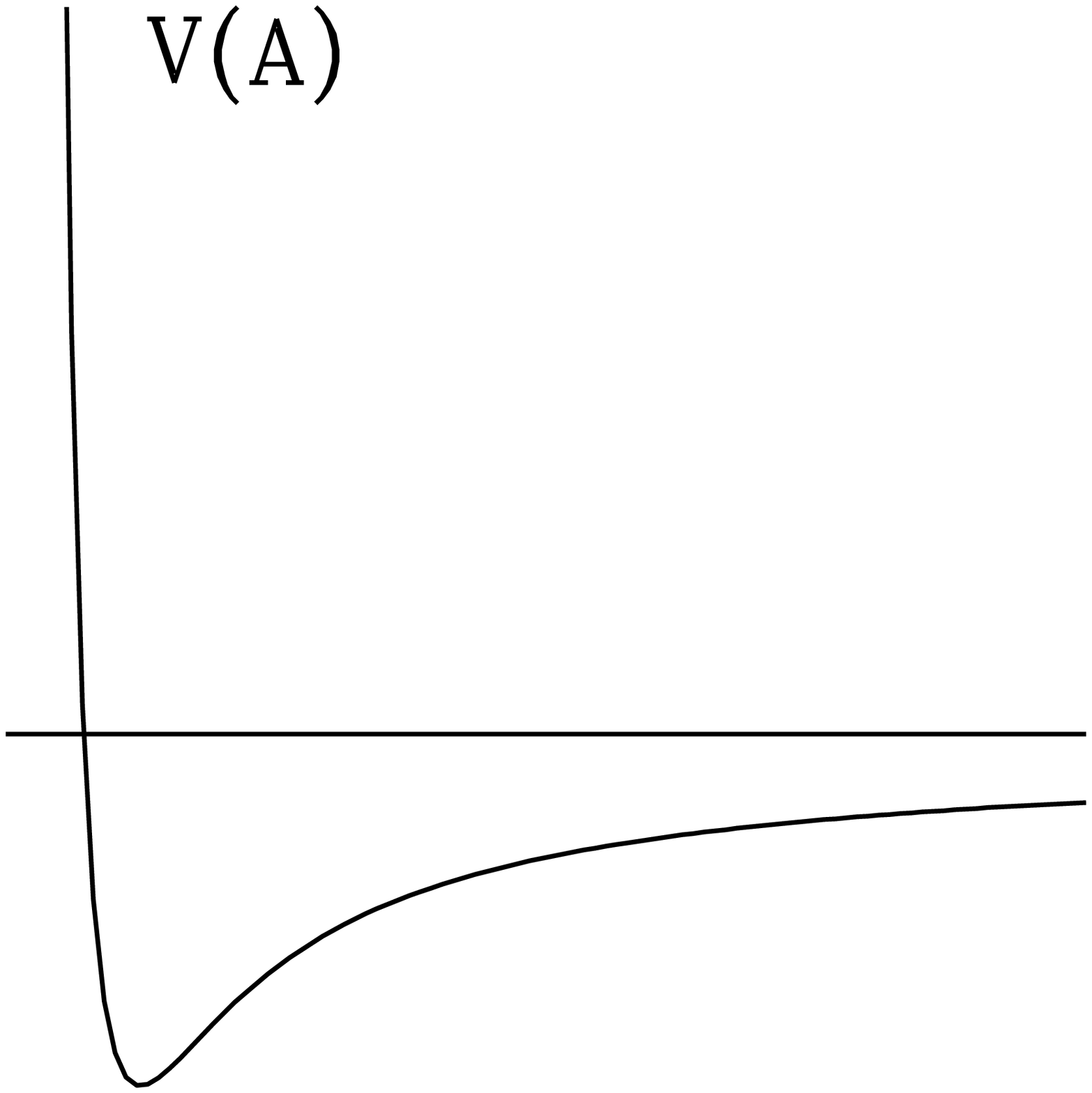}
{\bf (a)}\hspace{9cm}{\bf (b)}
\caption{\label{figure1} On the left (a), the shaded region indicates
the part of the $(\gamma,\xi)$ plane for which the potential $V(A)$
has the form displayed on the right (b), with the supplemental
condition $|w|\leq1$; the vertical line is related to the string frame
behavior (see paragraph \ref{substring}). The case $d=3$ has been
displayed, the other cases being qualitatively analogous.}
\end{center}
\end{figure}

Eq.~(\ref{acca}) can be rearranged in such a way that it has the
form
\beq\label{mech}
\frac{1}{2}\dot{A}^2+V(A)=0\, ,
\eeq
so that the evolution of the system can be easily studied as the
evolution of a point particle of unit mass and zero energy which moves
in the potential $V(A)$ given by
\beq\label{poten}
V(A)=\frac{1}{2 d^2}\left[C A^{p_1}-
\frac{\kappa^4}{16}\left(R_0^2 A^{p_2}+ 2 \lambda R_0
e^{\frac{\gamma}{2}\Phi_0}A^{p_3}\right)\right]\, ,
\eeq
with
\beq
p_1=-\frac{d \gamma^2}{2}-d +1\, ,\quad
p_2=-2d(1+w)-d\gamma^2 +2\, ,\quad
p_3=p_2 + d(1+w)\, .
\eeq
The parameters $\gamma$ and $\xi$ can be now chosen in such a way that
$V(A)$ has the appropriate form (shown in figure \ref{figure1}(b))
for producing the desired behavior, namely:
\begin{enumerate}
\item bounce:
\beq\label{potbounce}
V(A)>0\ {\rm as}\ A\rightarrow0^{+}\quad\Longrightarrow\quad\quad
p_1<p_2\, ,\quad C>0\, ,
\eeq
\item correct behavior at late time (decelerated expansion):
\beq\label{decel}
V(A)\rightarrow0^{-}\ {\rm as}\ A\rightarrow\infty\quad\Longrightarrow\quad
p_3<0\, ,
\eeq
\end{enumerate}
These conditions ensure also that at late times the
Friedmann equation is dominated by the term
$\sim \kappa^4 \rho_{\lambda}\rho$, thus recovering Einstein gravity
with an effective (dilaton-dependent) Newton constant proportional to
$\kappa^4 \rho_{\lambda}$.

As well as eqns.~(\ref{potbounce},\ref{decel}),
the condition $|w|\leq 1$ should also be imposed, thus giving
\beq\label{w<1}
(1-d)\leq\frac{\gamma}{2 \xi}\leq(1+d)\, .
\eeq
All the conditions (\ref{potbounce}), (\ref{decel}) and (\ref{w<1})
can be simultaneously satisfied, and the region in the $(\gamma,\xi)$
plane for which this happens is displayed in fig.~\ref{figure1}(a).

It can be shown that in the allowed region one has always $s<0$, which
in turn implies $\Lambda<0$ (this last condition coming from the fine
tuning relation $\frac{\kappa^4}{32}\lambda^2=\frac{\Lambda}{s}$).
Thus, the
region of parameter space of interest for the bounce is also the one
for which the bulk has a negative $\Lambda$, which is expected to
provide the confinement of a massless graviton mode on the brane.

It should also be noted that the conditions $C>0$, $s<0$ and $\Lambda<0$
imply the existence of a naked singularity in the bulk at $r=0$.
In the present context, this should not be regarded as a problem:
clearly, for small $r$ the equations of motion are not valid anymore
and should be replaced by a more complete description
(like string theory with loop and curvature corrections included),
which should be capable of damping the singularity,
or at least hiding it behind an horizon.
Though at the present stage we do not know how this mechanisms
work in detail, we will show in brief that there are solutions for which
the brane never comes close to the high curvature region: thus the
brane never exits the region of validity of General Relativity
(or low energy string theory) and its evolution can be studied without
knowing the details of the high curvature regime.

The evolution of the universe can be read directly in figure
\ref{figure1}(b): the scale factor starts very large, it then undergoes
a phase of accelerated contraction, which is an inflationary phase,
being characterized by $\dot{A}\ddot{A}>0$ (see \cite{infl}),
until it reaches the minimum of the potential.
Then the contraction rate slows down, until it becomes zero
(bounce) and a period of accelerated expansion starts (another
inflationary stage); finally, after having passed again through the
minimum of the potential, the brane enters a stage of decelerated expansion,
eventually dominated by the term $\sim G_N(\Phi) \rho$.
More precisely, at late times the Friedmann equation approaches the
following form:
\beq\label{latetimes}
H^2 \simeq \frac{\kappa^4}{16 d^2}2 \lambda R_0
e^{\frac{\gamma}{2}\Phi_0}A^{-d(1+w) -d\gamma^2}\, .
\eeq
Due to the running of the dilaton, the exponent of $A$ is not yet the one
which is found in standard cosmology. This means that, for instance,
a fluid with $w=1/d$ (radiation) would not induce the expected
behavior $\sim A^{-(d+1)}$ in the r.h.s. of eq.~(\ref{latetimes}).
As is typical in dilatonic models, this can be obtained only after
having achieved dilaton stabilization, a task which will not be
pursued here.

Incidentally, we notice that because of the running of the dilaton
it is possible to have a radiation-like behavior in the Friedmann equation
{\it without radiation}, i.e.~with $w\neq 1/d$: it is in fact
sufficient to tune $\gamma^2$ in such a way that the exponent of
$A$ in eq.~(\ref{latetimes}) is equal to $-(d+1)$. A quick calculation, and
comparison with eq.~(\ref{const}), shows that this
``ghost radiation'' scenario is realized for 
\beq\label{ghostr}
\gamma=\frac{1}{2 d \xi}\, ,
\eeq
and that the curve defined by this equation passes through the allowed
region in the $(\gamma,\xi)$ plane, meaning that this condition is
compatible with all the others that have been previously imposed.

\section{String frame behavior}\label{III}
\subsection{Pre-big bang model}
Let us now briefly recall the basic features of the graceful exit
problem in pre-big bang model.
The starting point is the minimal gravitational sector contained in
the low energy effective action of any critical 
superstring theory, which in the string frame has the following form:
\beq\label{sstring}
{\cal S}=\frac{1}{\lambda_{S}^{D-2}}
\int{\rm d}^{D}x \sqrt{g}e^{-\Phi}\left[{\cal R} +
\left(\nabla\Phi\right)^2\right]\, ,
\eeq
where $\lambda_{S}$ the string length, and $10-D$ spatial dimensions
are assumed to be compactified and frozen.
If one makes a flat cosmological ansatz, the equations of motion can
be written as:
\beq\label{pbb}
\left\{
\begin{array}{l}
\dot{H}=\pm\sqrt{D-1} H^2 \\
\dot{\Phi}=\left(D-1\pm\sqrt{D-1}\right) H\, . \\
\end{array}
\right.
\eeq
In the pre-big bang model, one has two disconnected branches of
solution: the first solves the equations of motion with the $(+)$
sign, has as initial conditions a Minkowski space with constant
dilaton, and describes a pre-big bang phase of superinflationary
expansion and growing dilaton (which means growing coupling constant);
the second, separated from the first by the big bang singularity,
solves the equations with the $(-)$ sign, and describes a phase of
decelerated expansion, which is supposed to be joined in a later time
to a radiation dominated phase with stabilized dilaton.

The graceful exit problem consists in joining these two branches and
thus ending with a single smooth evolution for the universe. Such a
problem can be studied also in the Einstein frame, i.e.~by conformally
transforming the metric in such a way that the action (\ref{sstring}),
expressed in terms of the new metric, describes a General Relativity
model minimally coupled to the dilaton, exactly like the bulk part of
eq.~(\ref{action}) (with $\Lambda$ set to zero).

The computation of the Einstein frame Hubble parameter is
straightforward
\beq\label{eframe}
g_{\mu\nu}^{E}=g_{\mu\nu}^{S}e^{-\frac{2 \Phi_{S}}{D-2}}
\Longrightarrow H_{E}=e^{-\frac{\Phi_{S}}{D-2}}\left[H_{S}-
\frac{\dot{\Phi}_{S}}{D-2}\right]\, ,
\eeq
where the exponential prefactor comes from the fact that the cosmic
times in the Einstein frame and in the string frame are related by
${\rm d}t_{E}=e^{\frac{\Phi_{S}}{D-2}}{\rm d}t_{s}$.
One can immediately check with the help of eqns.~(\ref{pbb})
that $H_{E}$ is negative in the pre-big bang phase and positive in the
post-big bang one, so that the graceful exit problem can be actually
mapped into that of having a bouncing universe in the Einstein frame.

In the pre-big bang model, this is achieved by considering the effect
of loop and high curvature corrections, which become
important as the solution approaches the singularity, and which can be
seen as effective NEC violating sources that provide the bounce.
As already mentioned in the introduction, here one encounters a
technical problem in that our knowledge of the exact form of these
corrections is still too limited.

\begin{figure}
\begin{center}
\includegraphics[width=5.0in, angle=0]{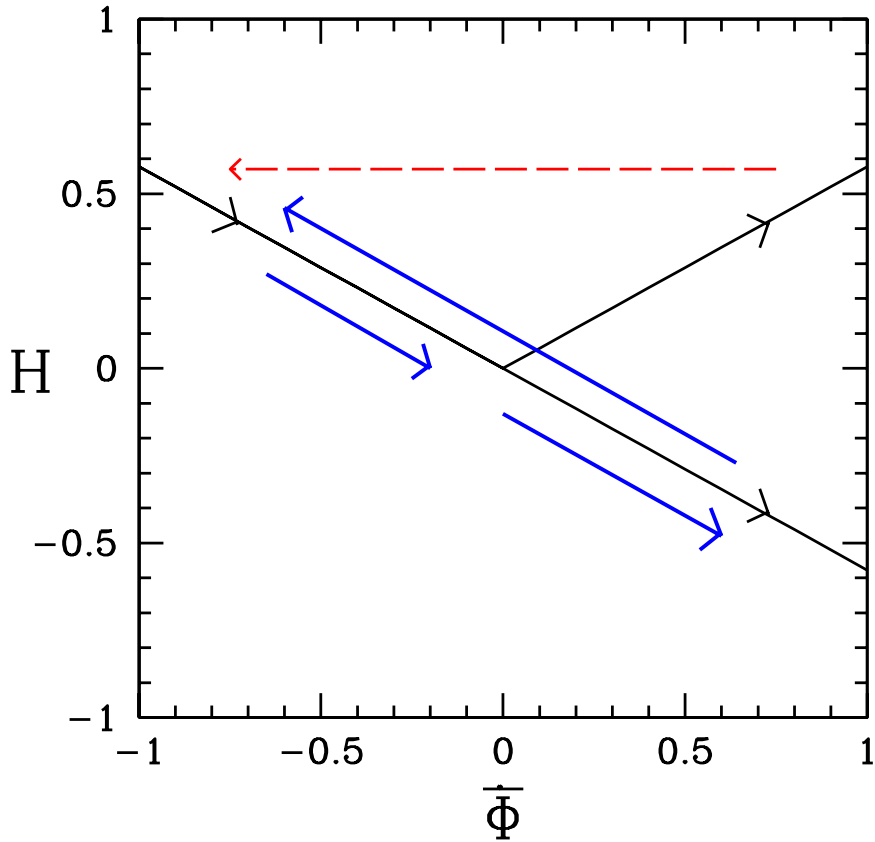}
\caption{\label{figure2} Different routes towards the exit in
the pre-big bang model, in the ekpyrotic one, and in the scenario
analyzed in the present work.}
\end{center}
\end{figure}

In string cosmology it is useful to represent the solution in the
$(\dot{\bar{\Phi}},H)$ plane, with
$\dot{\bar{\Phi}}\equiv\dot{\Phi}-dH$.
The origin of such a plane is Minkowski space with a constant dilaton
and represents, as well as the initial condition for the
pre-big bang model, also an attractor for the late time evolution.
With reference to figure \ref{figure2}, the pre- and post-big bang
branches are represented by the two lines in the upper part
of the plane, and realizing the graceful exit (dashed red arrow) consists in
connecting the pre-big bang line, which escapes from the origin, to
the post-big bang one.
The evolution described by the ekpyrotic scenario \cite{ekpy} can also
be represented in this plane: in this case the pre-big bang stage is
represented by the line in the lower part of the plane, and describes
a phase of accelerating contraction for the scale factor on the brane,
which corresponds also to the fact the two branes of the model are
approaching each other.
In the ekpyrotic model the problem of joining this line to the
post-big bang one is in some sense ignored, in that the scale factor
is let shrink to zero (the two branes collide as the evolution
proceeds along the pre-big bang line until infinity), and then starts
expanding again (the branes pass across each other, and the
evolution now proceeds from infinity towards the origin along the
post-big bang line).

\subsection{Comparison with the bouncing brane}\label{substring}
Although the variable $\dot{\bar{\Phi}}$ does not have a particular
dynamical meaning in the present context, the solution found in the
previous section can nevertheless be mapped in the same plane, in
order to conveniently compare it with the other models.
In order to do that, it is necessary to transform it in the string
frame, via
\beq\label{sframe}
A_{S}=A_{E}e^{\frac{\Phi_{E}}{\sqrt{2d}}}
=e^{\frac{\Phi_0}{2d}}A_{E}^{1-\gamma\sqrt{\frac{d}{2}}}\, .
\eeq
We now restrict our attention to the case
\beq\label{vline}
\gamma<\sqrt{\frac{2}{d}}
\eeq
in such a way that to an expanding Einstein frame scale factor, also
an expanding string frame one corresponds.
The condition (\ref{vline}) is satisfied at the left of the vertical
line in figure \ref{figure1}(a); note that $\gamma$ is still allowed
to be either positive or negative.

Given eqns.~(\ref{sframe}), the behavior of $\dot{\bar\Phi}_{S}$
can be found:
\beq
\Phi_{S}=\sqrt{\frac{d}{2}}\Phi_{E}\quad\Rightarrow\quad
\dot{\bar\Phi}_{S}=-\frac{d}{1-\gamma\sqrt{\frac{d}{2}}}H_{S}
\quad\left(\Rightarrow\frac{\dot{\bar\Phi}_{S}}{H_{S}}=const.<0
\right)\, .
\eeq
Thus, as in the ekpyrotic case, the solution still moves along a line
of negative slope in the $(\dot{\bar{\Phi}},H)$  plane, but in this
case it follows a somehow less drastic route to the exit, simply
reversing at some point its motion along this line, passing
trough the origin (bounce), and then being smoothly connected to the
decelerated expanding phase. Such a behavior is represented by the
blue arrows in figure \ref{figure2}. 

Differently from what happens in the standard pre-big bang case, now
the exit can be realized at low curvature and low coupling.
To ensure the first point, it is sufficient that the free parameter
$C$ is big enough; as to the second request, one can take $\gamma<0$,
so that (as eq.~(\ref{charm}) tells) if the coupling is small at large
$A$ (certainly a desirable feature, if one wants this model to be
connected to our universe at late time), then it will be even
smaller during the bouncing phase.
Thus, the whole evolution takes place in a regime under full
theoretical control.

\section{Comments and conclusions}
The solution discussed in this work is an example of
bouncing universe in the context of Einstein brane gravity coupled
to a scalar field.
When transformed in the string frame, such a solution describes a
pre-big bang model where the transition between pre- and post-big bang
is realized at low curvature and weak coupling, thus providing
an example of graceful exit which takes place under complete
theoretical control.

The investigation focused on the graceful exit aspect, while
other issues, such as the naturalness of the initial
conditions, the amount of inflation, the dilaton stabilization,
have not been addressed, partly because this goes beyond the purposes
of this paper, and partly for a more technical reason.
In fact, while from the discussion of paragraph \ref{genweyl} it should
be clear that the bouncing mechanism is not related to the particular
kind of bulk background chosen for the present analysis
(depending only on the sign of the bulk Weyl tensor), on the other
hand the same can not be said about other important features of the
cosmological behavior.

The quantity of inflation produced in this model certainly depends
on the form of the potential $V(A)$, and thus on the exact form of the
bulk background. Also the initial configuration does, and actually one
could argue that a brane which starts falling towards a naked
singularity does not look natural at all; this problem could be
addressed by moving to more general, time dependent bulk backgrounds,
maybe like the one discussed in \cite{BDV} in the context of initial
conditions for the pre-big bang model.

On the other hand, the background solution chosen in this work
is simple enough for emphasizing the main point of this
paper, i.e.~that in brane world models it is possible to realize
a graceful exit without resorting either to quantum gravity
(or stringy) corrections, or to exotic forms of matter.
The embedding of the mechanism studied in the present paper into
more complete cosmological scenarios is left for future
investigations.

\end{document}